\documentstyle[aps,prl,multicol,epsf]{revtex}
\begin{document}       
\title{\bf A Nonconservative Earthquake Model of Self-Organized
           Criticality on a Random Graph} \author{Stefano Lise and
           Maya Paczuski}
 
\address{Department of Mathematics, Huxley Building, Imperial College
of Science, Technology, and Medicine, London UK SW7 2BZ \\}
\date{\today}
 
\maketitle 
 
\begin{abstract}
We numerically investigate the Olami-Feder-Christensen model on a
quenched random graph. Contrary to the case of annealed random
neighbors, we find that the quenched model exhibits self-organized
criticality deep within the nonconservative regime. The probability
distribution for avalanche size obeys finite size scaling, with
universal critical exponents. In addition, a power law relation
between the size and the duration of an avalanche exists. 
We propose that this may represent the correct mean-field limit of the model
rather than the annealed random neighbor version.
\end{abstract}
 
\vspace{0.3cm}

\begin{multicols}{2}

The idea of self-organized criticality (SOC) was introduced as a
possible explanation for the widespread occurrence in nature of long
range correlations in space and time \cite{BTW}. The term refers to
the intrinsic tendency of a large class of spatially extended
dynamical systems to spontaneously organize into a dynamical critical
state.  In general, SOC systems are driven externally at a very slow
rate and relax with bursts of activity, avalanches, on a very fast
time scale. One signature of SOC is a scale free, e.g. power law,
distribution of avalanche sizes.   This is normally related to some
long range spatial and temporal correlations within the system.
Typical natural realizations of this phenomena include, among others,
earthquakes, forest fires, and biological evolution (for 
reviews, see \cite{bak_book,jen_book}).

A problem that has attracted a lot
of attention, but is still poorly understood, is that of identifying
fundamental mechanisms leading to SOC behavior. In particular, much
effort has been directed at understanding how
conservation of the transported quantity (e.g. sand) in the avalanche
dynamics affects criticality~\cite{hwa-kardar,grinstein}. For
instance, it is well known that the Abelian sandpile model \cite{BTW},
is subcritical if dissipation is introduced~\cite{manna_1}.  On the
other hand, nonconservative sandpile models which display criticality
have since been introduced~\cite{ali,manna_2}. Although both the
analytical and numerical evidence in favor of criticality are quite
convincing, the role played in these models by the non-conservative
dynamics is not clear.  In fact, dissipation is a dynamical variable
and does not always occur.

A model, which in the context of SOC in nonconservative systems has played an 
important role, is the Olami-Feder-Christensen (OFC) model of
earthquakes~\cite{ofc}.  In the OFC model a finite fraction,
controlled by a fixed parameter $\alpha$, of the transported
quantity is dissipated in each relaxation event.  The presence of
criticality in the non-conservative OFC model has been controversial
since the very introduction of the model~\cite{klein} and it is still
debated~\cite{carvalho,kim}.  Recent numerical investigations, though,
have shown that the OFC model on a square lattice displays scaling
behavior,  up to lattice sizes presently accessible by
computer simulations~\cite{lisepac1,lisepac2}.  The avalanche size
distribution is described by a power law, characterized by
a universal exponent $\tau \simeq 1.8$, independent of the dissipation
parameter.  This distribution does not display finite size scaling, however.

To overcome the limitation of relying almost exclusively on computer
simulation results, it has sometimes been useful to consider an
annealed random neighbor (RN) version of the
model~\cite{lise,chabanol,broker,kinouchi}, where each site interacts
with randomly chosen sites instead of its nearest neighbors on the
lattice. This considerably simplifies the problem. In the past, RN
models have usually been considered as mean-field descriptions of
their fixed lattice counterparts, since spatial correlations are
absent. Analogous to other RN models, the RN OFC model can be solved
analytically~\cite{chabanol,broker}. It displays criticality only in
the conservative case, where it becomes equivalent to a critical
branching process. As soon as some dissipation is introduced the
avalanches become localized, although the mean avalanche size diverges
exponentially as dissipation tends to zero.  The absence of
criticality, together with the exponential divergence has cast some
doubt on whether the OFC model on a fixed lattice is critical.

However, it is important to point out that the RN model may not
describe the behavior of the OFC model on a fixed lattice in any
dimension, and thus may not correspond to the mean field limit of the
model.  Usually, mean field behavior describes the high dimensional
behavior of the system (e.g. the behavior above an upper critical
dimension); this is not exactly the same limit as a model without any
spatial correlations in it.

In fact, criticality in the OFC model on a lattice has been ascribed
to a mechanism of partial synchronization~\cite{middleton}. The system
has a tendency to order into a periodic
state~\cite{middleton,socolar,grass2} which is frustrated by the
presence of inhomogeneities such as the boundaries. In addition,
inhomogeneities induce partial synchronization of the elements of the
system building up long range spatial correlations and thereby
creating a critical state.  The mechanism of synchronization requires
an underlying spatial structure and therefore cannot operate in an
annealed RN model, where each site is assigned new random neighbors at
each update.

The main purpose of the present work is the investigation of the OFC
model on a quenched random graph. This formulation, which can be
handled numerically, is worth analyzing to see if it presents critical
or non-critical behavior.  Since the largest distance between two
sites in the random graph grows only as a logarithm of the number of
sites, it can be considered to be a high dimensional limit of a
lattice model, and thus may describe the correct mean field limit.
 Contrary to the RN case, in a random graph the choice
of neighbors is not annealed but quenched,  so that 
spatial correlations can develop.  Indeed, we  show that
the OFC model on a random graph displays criticality even in the
nonconservative regime.

A random graph is defined as a set of $N$ sites connected at random by
bonds.  Two connected sites are denoted as
``nearest-neighbor''. Formally, the random graph can be constructed by
considering all $N(N-1)/2$ possible bonds between sites and occupying
a certain number of them with equal probability. A constraint of fixed
connectivity can also be imposed by requiring that each site has the
same number of neighbors, $q$. We have mainly concentrated on this
latter situation but the first case will also be discussed.  The model
is then defined as follows.  To each site of the graph is associated a
real variable $F_i$, which initially takes some random value in the
interval $(0,F_{th})$.  All the forces are increased uniformly and
simultaneously at the same speed, until one of them reaches the
threshold value $F_{th}$ and becomes unstable $(F_i \geq F_{th})$. The
uniform driving is then stopped and an ``earthquake'' (or avalanche)
starts:
\begin{equation}
 \label{av_dyn} 
           F_i \geq F_{th}  \Rightarrow \left\{ \begin{array}{l}
                                       F_i \rightarrow 0 \\
                         F_{nn} \rightarrow F_{nn} + \alpha F_i
                                      \end{array} \right.             
               \end{equation}                       
where 
``nn'' denotes the set of nearest-neighbor sites of $i$. The parameter $\alpha$
controls the level of conservation of the dynamics and, in the case of a
graph with fixed connectivity $q$, it takes values between $0$ and $1/q$ 
($\alpha=1/q$ corresponding to the conservative case).
The toppling rule (\ref{av_dyn}) can possibly create new unstable sites, 
producing a chain reaction. All sites that are above threshold at a 
given time step in the avalanche relax simultaneously according to 
(\ref{av_dyn}) and the earthquake is over when there are no more unstable 
sites in the system ($F_i < F_{th}$, $\forall i$). The uniform growth then 
starts again. The number of topplings during an earthquake defines its size, 
$s$, and we will be interested in the probability distribution $P_N (s)$. 
Another quantity of interest is the duration $t$ of an earthquake which will 
be identified with the number of time steps needed for the earthquake to 
finish.

We consider first a random graph where all sites have exactly the same number
of nearest neighbors $q$. In this case, we have verified (both for $q=4$ and
$q=6$) that the system organizes into a subcritical state. This is analogous
to what happens in the OFC model on a lattice with periodic boundary 
conditions, where no critical behavior is 
observed~\cite{middleton,socolar,grass2}.  
In order to observe scaling in the avalanche distribution, one has to 
introduce some inhomogeneities. In the lattice model this is generally 
achieved by considering open boundary conditions which imply that boundary 
sites  have fewer neighbors and therefore cycle at a different frequency 
from bulk sites. This is an inhomogeneity with a diverging length scale in
the thermodynamic limit. For the OFC model on a random graph, we have 
found that it suffices to consider just two sites in the system with 
coordination $q-1$~\cite{note1}. When either of
 these sites topple according to rule 
(\ref{av_dyn}), an extra amount $ \alpha F_i$ is simply lost by the system.

After a sufficiently long transient time, the system settles into a
statistically stationary state. We have verified that the statistical
properties of the system (e.g. the avalanche distribution) are
independent of the actual realization of the random graph, as long as
the coordination number $q$ is the same. As a point of comparison, in
figure~1 we report the probability distribution of avalanche sizes for
(a) the annealed RN model and (b) for the OFC model on a random graph
for various system sizes $N$.  The dynamical rule for the annealed RN
model are formally similar to (1), where, instead of the
nearest-neighbor sites on the graph, $q$ new random sites are chosen
at each relaxing event.  In both cases of fig.~1, the number of
neighbors is $q=4$ and the parameter $\alpha=0.10$. It is clear that
no scaling is present in the RN model as the cut-off in the avalanche
size distribution does not grow with system size.  On the contrary for
the model on a random graph, the distribution scales with system size,
which is indicative of a critical state.  In fact, the largest
avalanche roughly coincides with system size.  It is important to
underline that we are considering a situation far away from the
conservative case (60\% of the force in the toppling site is
dissipated) and therefore one could expect that if a finite length
scale related to conservation existed in the system it should appear
for system sizes we have considered.

In order to characterize the critical behavior of the model,
a finite size scaling (FSS) ansatz is used, i.e.
\begin{equation}
\label{fss}
P_N(s) \simeq N^{-\beta} f(s/N^D)
\end{equation}
where $f$ is a suitable scaling function and $\beta$ and $D$ are
critical exponents describing the scaling of the distribution
function.  In figure~2, a FSS collapse of $P_N(s)$ for different
values of $\alpha$ and for different $q$ is shown. The distribution
$P_N(s)$ satisfies the FSS hypothesis reasonably well, with universal
critical coefficients.  The critical exponent derived from the fit of
fig.~2 are $\beta \simeq 1.65$ and $D=1$, independent of the
dissipation parameter $\alpha$ and the coordination number of the
graph $q$. The FSS hypothesis implies that, for asymptotically large
$N$, $P_N(s) \sim s^{-\tau}$ and the value of the exponent is $\tau=
\beta/D \simeq 1.65$. Due to the numerical uncertainty on the estimate
it is difficult to assert with certainty
that $\tau$ is a novel exponent, different
from the one for the conservative
RN model ($\tau=1.5$) or the lattice model in two dimensions
($\tau \simeq 1.8$).

The OFC model on a lattice does not show ordinary FSS~\cite{grass2}. Although
the avalanche size distribution converges to a well-defined, universal power
law, the cut-off in the distribution due to finite system sizes does not 
behave according to FSS~\cite{lisepac1}. In particular, the apparent numerical
value  for the exponent $D$ determined through FSS would violate some exact
bounds~\cite{klein}. In fact, in the non-conservative model each site can
only discharge a finite number of times during an avalanche, which  
imposes $D \le 1$ in  eq.~(\ref{fss}). In order to recover standard FSS
in the two dimensional lattice model one
has to consider earthquakes localized within subsystems of linear dimension
small compared to the overall system size~\cite{lisepac2}. 

We have performed numerical simulations of the OFC model on a square two 
dimensional periodic lattice where just two sites have $3$ neighbors. We 
have verified that the avalanche size distribution scales with system size 
but, as for the OFC model with open boundary conditions, FSS appears to be 
violated in the cut-off region. The power law exponent of the
distribution is consistent with the exponent of the OFC model with open 
boundary conditions, i.e. $\tau \simeq 1.8$. For the range of system sizes 
we could simulate, the critical behavior of the model on a lattice and on a
random graph (with the same number of defects) appear to be different.
In particular SOC on the quenched random graph appears to be described
by ordinary FSS, consistent with a  mean field limit.

We now discuss the time properties of the avalanches. In 
fig.~3 we report the average size of an avalanche stopping at time $t$,
$<s>_t$, as a function of the rescaled time $\tilde{t}=t+2$ (as we are mainly 
interested at large values of $t$, the constant should be irrelevant).
The curves for different system sizes overlap (deviations can be attributed
to finite-size effects) and we observe that $<s>_t \simeq t ^{\gamma}$, where 
$\gamma \simeq 2.1$, providing  further evidence of criticality in 
the nonconservative system.

An interesting question (of difficult solution though) is whether the model 
becomes subcritical below a certain non-zero value $\alpha _c$ (for 
$\alpha = 0$ the system is clearly not critical as sites do not interact). 
In our simulations we have found that for $\alpha \ge 0.10$ and $q=4$ the 
model displays scaling behavior with universal critical exponents 
(see fig.~1 and 2). For lower values of $\alpha$ the analysis is more 
complicated as the extremely long transient times to stationarity prevent the 
investigation of large lattices. Nonetheless for very low values of $\alpha$ 
($\alpha \simeq 0.03$) the cut-off in the avalanche distribution does not 
appear to vary systematically with system size (even for relatively small 
systems), suggesting that a non-zero $\alpha _c$ may exist.
 

We have also considered the OFC model on a random 
graph with variable local connectivity $q_i$. In this case, the toppling
rule (\ref{av_dyn}) must be modified to take into account that different 
sites have a different coordination number $q_i$. Each site consequently 
has a different $\alpha _i$, which we determined by requiring  that the 
total fraction $\tilde{\alpha}$ of the force transferred 
from the unstable site to the nearest-neighbors sites is constant in the 
system, i.e. $\alpha _i=\tilde{\alpha}/q_i$. In particular, we have 
studied a graph with average connectivity $<q_i> = 4$. We have found that 
there is no criticality in the system as the cut-off in the probability 
distribution does not scale with system size.  In agreement with previous 
investigations~\cite{ceva,mousseau}, this result indicates that
if the disorder is too strong (as for a completely random graph) the critical
state is destroyed. On the other hand inhomogeneities are necessary to break
the periodic state the system would otherwise reach. It is a difficult 
question to establish what is the maximum level of disorder that the system
can sustain without loosing its critical properties.


In conclusion, in this paper we have investigated the OFC model on a quenched
random graph. We have shown that the model is critical even in the 
nonconservative regime.
This is in contrast to what happens in the annealed
RN OFC model which displays criticality only in the conservative case. 
Contrary to the annealed case, a quenched random graph has an underlying 
spatial structure so that partial synchronization of the elements of the 
system can still occur. As a random graph can be regarded as a high 
dimensional limit of a regular lattice, we propose that this formulation 
represents the correct mean-field limit of the model rather than the 
annealed random neighbor version.

\medskip

This work was supported by the EPSRC (UK),  
Grant No. GR/M10823/01 and Grant No. GR/R37357/01.

\begin{figure}[hb]
\narrowtext
\epsfxsize=4in
\centerline{\epsffile{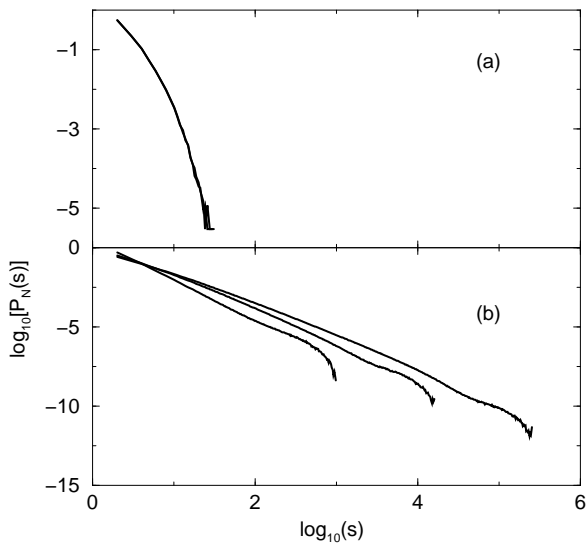}}
\protect\vspace{0.2cm} 
\caption[1]{\label{fig_1a}
Probability distribution (a) for the RN OFC model and (b) for the 
OFC model on a random graph. In both cases, $q=4$ and $\alpha=0.10$. 
System sizes are (a) $N=10^3$, $4\cdot 10^3$ and (b) $N=10^3$, $16 \cdot 10^3$,
$256 \cdot 10^3$.
}
\end{figure}

\begin{figure}[hb]
\narrowtext
\epsfxsize=4in
\centerline{\epsffile{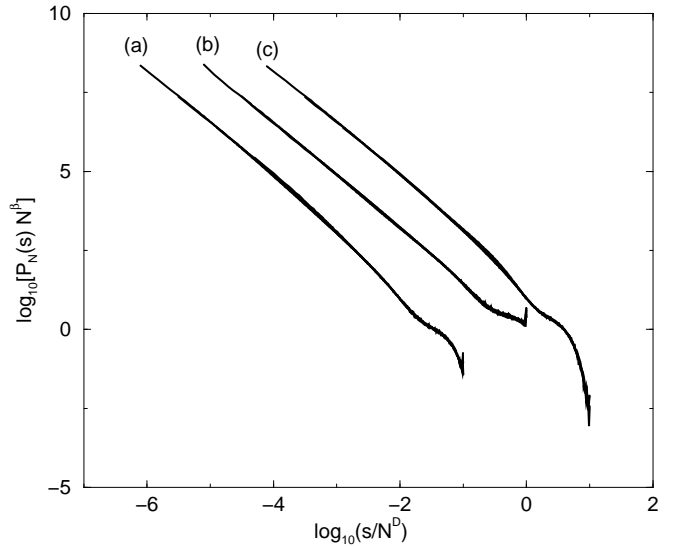}}
\protect\vspace{0.2cm} 
\caption[1]{\label{fig_2}
Finite-size scaling plots for $P_N(s)$ for (a) $q=4$, $\alpha=0.15$,  
(b) $q=4$, $\alpha=0.20$ and (c) $q=6$, $\alpha=0.10$. System sizes are
$N=4\cdot 10^3$, $16 \cdot 10^3$, $64 \cdot 10^3$ and $256 \cdot 10^3$.
The critical exponents are $\beta=1.65$ and $D=1$.
For visual clarity, curves (a) and (c) have been shifted along the x axis,
$x \rightarrow x-1$ and $x \rightarrow x+1$, respectively.
}
\end{figure}

\begin{figure}[hb]
\narrowtext
\epsfxsize=4in
\centerline{\epsffile{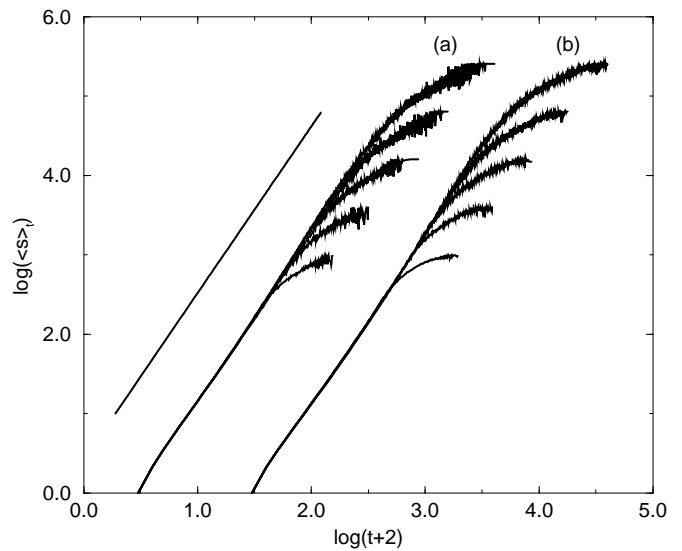}}
\protect\vspace{0.2cm} 
\caption[1]{\label{fig_3}
Average size of an avalanche lasting $t$ time steps as a function of $t$ for
$q=4$ and (a) $\alpha=0.15$ and (b)  $\alpha=0.20$. Different curves 
correspond, from bottom to top, to system sizes $N=1\cdot 10^3$, 
$4\cdot 10^3$, $16 \cdot 10^3$, $64 \cdot 10^3$ and $256 \cdot 10^3$.
The slope of the straight line is $\gamma =2.1$. Curve (b) has been shifted 
along the $x$ axis, $x=x+1$, for visual clarity.
}
\end{figure}

\end{multicols}

\begin{thebibliography}{99}
\bibitem{note1}
We have also considered defects whose number scale as $\sqrt{N}$ (analogously 
to boundary sites in the lattice model). The model still displays scaling but 
the results appears easier to interpret with just two defects.
\end{thebibliography}
\end{document}